%% file: main.tex
\documentclass{article}

\usepackage{PRIMEarxiv}

\usepackage[utf8]{inputenc} % allow utf-8 input
\usepackage[T1]{fontenc}    % use 8-bit T1 fonts
\usepackage{hyperref}       % hyperlinks
\usepackage{url}            % simple URL typesetting
\usepackage{booktabs}       % professional-quality tables
\usepackage{amsfonts}       % blackboard math symbols
\usepackage{nicefrac}       % compact symbols for 1/2, etc.
\usepackage{microtype}      % microtypography
\usepackage{lipsum}
\usepackage{fancyhdr}       % header
\usepackage{graphicx}       % graphics
\usepackage{subfig}
\usepackage{float}
\graphicspath{{media/}}     % organize your images and other figures under media/ folder

\title{Skill Profiling with Attributable Reasoning (SPAR): A Wearable Analysis System for Boxing}
\author{
  \textbf{Nibraas Khan}$^{1}$\thanks{Corresponding author: \texttt{nibraas.a.khan@vanderbilt.edu}} \quad
  \textbf{Hanchen David Wang}$^{1}$ \quad
  \textbf{Enya Bullard}$^{1}$ \quad
  \textbf{Ritam Ghosh}$^{2}$ \\
  \textbf{Ruj Haan}$^{4}$ \quad
  \textbf{Aarav Agrawal}$^{5}$ \quad
  \textbf{Meiyi Ma}$^{1}$ \quad
  \textbf{Nilanjan Sarkar}$^{1,3}$ \\[1em]
  \small $^{1}$Department of Computer Science, Vanderbilt University, Nashville, TN 37235, USA \\
  \small $^{2}$Department of Electrical Engineering, Vanderbilt University, Nashville, TN 37235, USA \\
  \small $^{3}$Department of Mechanical Engineering, Vanderbilt University, Nashville, TN 37235, USA \\
  \small $^{4}$Independent Researcher \\
  \small $^{5}$Milton Academy, Milton, MA 02186, USA
}

\begin{document}
\maketitle

\begin{abstract}
A punch is a ballistic, full-body action driven by a kinetic chain running from the legs through the trunk to the arm, where a small sequencing error separates a scoring strike from a miss. Wearable sensors can capture this movement in the gym, but most deployable systems only classify which punch was thrown rather than assess how well it was thrown. We present Skill Profiling with Attributable Reasoning (SPAR), an eight-IMU garment and pressure-insole system that classifies each punch as expert or novice and treats an explanation of that prediction as feedback. Feedback is only useful if the person receiving it can act on it, so SPAR explains the prediction at three tiers, a per-joint attribution for the analyst, a counterfactual over kinetic-chain layers for the coach, and a plain-language narrative of the two for the athlete. Across 17 participants and 4{,}713 punches, SPAR reaches a leave-one-participant-out AUC of $0.842$ (95\% CI $[0.769, 0.907]$ over participants). A frozen time-series foundation model encodes the joint-angle and plantar-force series, and a small transformer trained on the cohort classifies the encoding. We audit the two quantitative tiers and report six themes from a thematic analysis of interviews with six practicing boxing coaches.
\end{abstract}

% keywords can be removed

\begin{figure}[t]
  \centering
  \includegraphics[width=\textwidth,keepaspectratio]{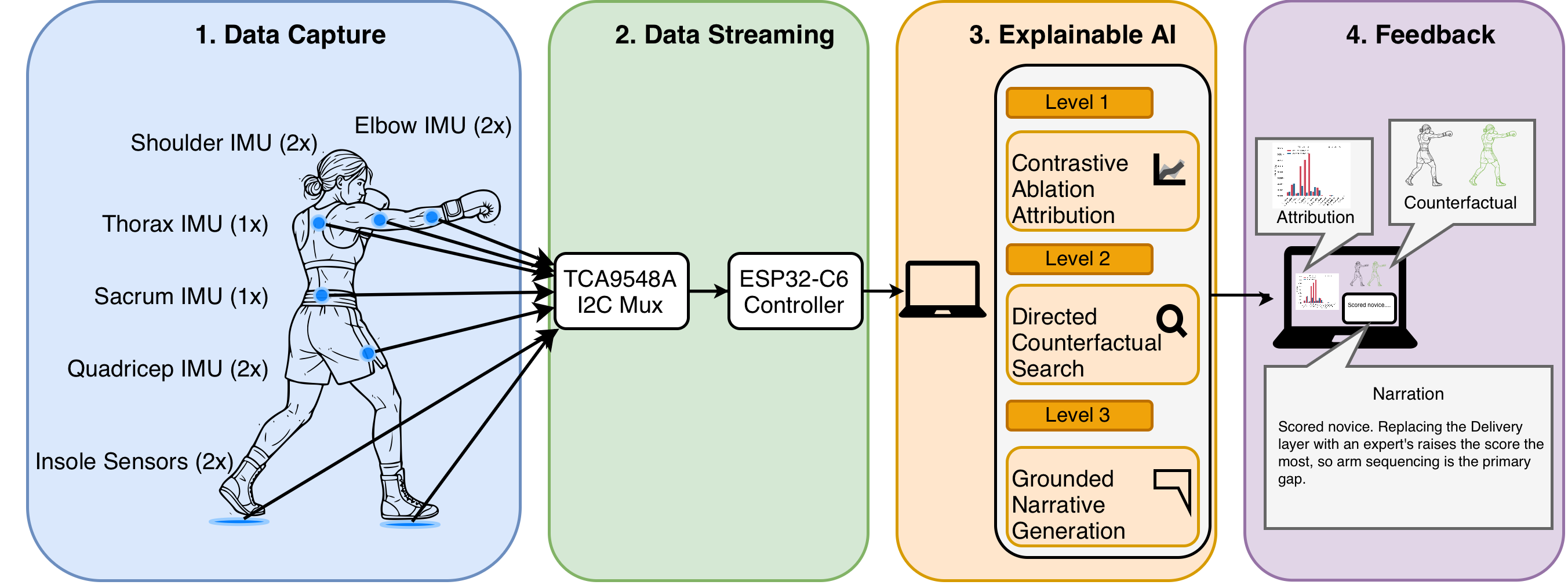}
  \caption{End-to-end overview. An eight-IMU garment and pressure insoles stream to a skill classifier over anatomical joint angles. Three explanation tiers stack on the per-punch skill prediction.}
  \label{fig:overview}
\end{figure}

\section{Introduction}

Sport is a complex, high-stakes activity in which performance depends on fine margins. Both the outcome of a contest and the safety of the athlete often depend on details too fast and too fine for the eye to resolve. These details are the segmental timing, coordination, and force sequencing of a movement that happens over a few hundred milliseconds. Athletes improve by analyzing and refining these details of execution. Motion analysis has become increasingly data-driven, with coaches and sport scientists measuring movement quantitatively rather than judging it by eye alone \cite{camomilla2018trends, souaifi2025artificial}, as visual analysis is too coarse to resolve fine details or to track how they shift over time \cite{ressman2021visual}.

This is especially true of a ballistic sport such as boxing. Punches generate forces large enough that head impact is a documented injury mechanism \cite{walilko2005biomechanics}, and how much force a punch carries depends on technique~\cite{lenetsky2013assessment}. Coaches typically assess technique through observing an athlete, understanding their form, and providing cues based on subjective past knowledge that is often difficult to articulate or pass to another coach \cite{lindsay2020contribution}. It also does not scale. In a group class the same assessment has to cover several athletes at once, and individual diagnosis becomes nearly impossible. What the coach is doing in that moment is skill assessment. That judgment is what makes feedback actionable. An assessment of execution identifies areas for improvement, and in boxing, this execution comes down to how the athlete sequences their movements.

That sequencing is the proximal-to-distal coordination of the kinetic chain. Force begins in the legs, passes through the trunk, and the arm delivers it, an order that unfolds in a few hundred milliseconds~\cite{lenetsky2013assessment, dinu2020biomechanical}. Boxing is built around six canonical punches, the straight jab and cross, the lead and rear hook, and the lead and rear uppercut. Each one drives that chain differently. Despite this complexity, most existing work in boxing has focused on the simple question of which punch was thrown. These systems track movement at the wrist or upper limb and classify the strike type with high accuracy~\cite{worsey2020evaluation, manoharan2023punch, xue2025limb}. A smaller line of work reports physical outcomes such as punch force and velocity~\cite{menzel2021application}, or pairs a glove-mounted sensor with pressure insoles to separate professional from amateur boxers~\cite{sha2026wearable}. While these systems give an athlete useful information, improving technique requires a system that judges skill and turns that judgment into a cue.

Any such judgment rests on measurements of the movement itself, and those measurements have so far come mostly from laboratory studies of punch mechanics~\cite{tajibaev2024forward, dinu2020biomechanical, stewart2025role, liu2022biomechanics}. That work achieves a fidelity no field method matches, but it is not where feedback has to be delivered. A boxer who throws isolated punches inside a capture volume while wearing markers is not moving the way they would in a real training session. The faults worth correcting are the ones that emerge under the strain and fatigue of real training. Measuring how well a punch is thrown therefore requires capturing the whole kinetic chain at the speed it operates during ordinary training rather than in a laboratory.

Sensors worn on the body can capture movement under these conditions. Because they travel with the athlete, there is no capture volume to work inside and no camera angle to stay within, which makes the measurement ecologically valid in a way that laboratory instrumentation is not. The punch-classification systems described earlier already rely on this kind of sensing, though mostly from a single wrist sensor. We extend that sensing across the whole kinetic chain, and with it the scope of wearable technology in boxing from punch detection to skill assessment. We do this through our system, Skill Profiling with Attributable Reasoning (SPAR), which classifies each punch as expert or novice, uses that prediction as a proxy for skill, and explains what drove the difference. SPAR draws on four lines of research: movement sensing, deep learning, boxing biomechanics, and explainable AI.

\emph{Movement measurement from laboratory to field.} Other than manual observation, video review is the most common tool in coaching practice. Athletes record themselves performing a movement and review the footage to see what went wrong in their form and how to improve. This analysis is qualitative, bounded by camera angle and the reviewer's eye, and assessments vary depending on who reviews the footage \cite{ressman2021visual}. For more quantitative measurement, athletes turn to optical motion capture where depth cameras, force plates, and other high-precision tools can recover joint kinematics and ground reaction forces at a level of detail that the eye and video cannot reach \cite{meister2011rotational, diffendaffer2023clinician}. However, this instrumentation is constrained the lab and cannot capture how an athlete moves under the ecological demands of real training. Markerless vision extends capture to less constrained settings \cite{kanko2024comparison} where just a camera can capture kinematics. This is more ecologically valid, but it comes with its own constraints such as occlusion, limited field-of-view, and privacy concerns. 

Body-worn sensors avoid these limits by removing the need for an external recording \cite{camomilla2018trends}. Placed on the athlete, they can capture signals in any setting rather than only in a laboratory. Inertial Measurement Units (IMUs) capture segment orientation and acceleration, pressure insoles record plantar force, surface electromyography registers muscle activation \cite{marimon2024kinematic, burns2019validation, xue2025limb}. The IMU is one of the most common wearable sensors. It is small, inexpensive, low-power, and versatile enough to capture various movements such as the cyclical strokes of swimming and the ballistic punches of boxing.

Researchers have applied wearable sensing across many sports. In team sports, microsensors quantify locomotor load and detect sport-specific movements on the field \cite{chambers2015use}. In swimming, one study used six body-worn IMUs to segment laps, identify the stroke technique, and time the phases within each lap over a full training session \cite{hamidi2021novel}. In baseball, a forearm-worn IMU estimates elbow torque and arm speed for pitching-workload monitoring and has been validated against marker-based capture \cite{camp2021wearable}. Across these systems the sensor set stays small. Field IMU studies typically place one to six sensors on the segment of interest, sample between 50 and 250\,Hz, and tie their analysis to one sport's movements \cite{camomilla2018trends, rana2020wearable, souaifi2025artificial}. Clinical IMU work is most mature in gait analysis, where two sensors per leg recover joint angles with errors close to optical capture \cite{seel2014imu}. 

Several wearable systems also pair the sensors with a biomechanical model that fits a skeleton to the body-worn orientations and recovers the joint angles a laboratory would report~\cite{delp2007opensim, al2022opensense}, among them real-time full-body kinematics from IMUs and a microcontroller~\cite{slade2021open}, gait kinematics and joint loads outside the laboratory~\cite{karatsidis2019musculoskeletal}, spine kinematics during walking and running~\cite{sibson2024using}, and knee kinematics during soccer kicks~\cite{xu2025knee}.

\emph{Deep learning on wearable signals.} Regardless of the sensor, turning its signal into a decision about the movement requires an analytical layer that activity-recognition research has refined over more than two decades \cite{lara2012survey}. The earliest analyses relied on simple summary statistics of the raw signal. Machine learning soon took over, but for years it relied on the same two-stage pipeline. First, reduce a window of the signal to a set of hand-crafted features, then pass those features to a classifier \cite{bulling2014tutorial}. One early study extracted time- and frequency-domain features such as mean, energy, spectral entropy, and correlation from body-worn accelerometers and recognized everyday activities with a decision tree \cite{bao2004activity}. Later work compared large banks of such features, spanning time-domain, frequency-domain, and wavelet descriptors, to find which best separate dynamic activities \cite{preece2008comparison}. The classifiers evolved alongside the features, from decision trees and k-nearest neighbors to support vector machines and random forests \cite{lara2012survey}. What did not change was the reliance on features a human had to design.

Deep learning removed the hand-design step by learning features directly from the raw signal. The dominant architectures each capture a different property of how a movement unfolds in time. First, a movement is made of short local patterns, and any one of them can occur at any time within it. Convolutional neural networks capture these patterns with local, shift-invariant filters that respond whenever a pattern occurs \cite{yang2015deep, ignatov2018real}. Second, a movement is a sequence in which each timestep follows the one before. Recurrent neural networks model this dependence, carrying a hidden state forward so that each timestep builds on those before it \cite{ordonez2016deep, nweke2018deep}. Third, a model can link phases far apart in a movement, as when early leg drive sets up a later arm strike. Transformers capture these links, using self-attention to connect any two timesteps directly rather than passing the link step by step through a recurrent state \cite{dirgova2022wearable}. These models are now the default for human activity recognition.

These models are usually trained as supervised classifiers, which requires a large labeled dataset. Building that dataset is expensive in part because humans need to decide what the data represents and provide a label. To reduce this cost, the field has turned to self-supervised pretraining. A model first learns the structure of the raw signal on its own, with no labels, by reconstructing it \cite{vincent2010stacked}. A downstream task then builds on this learned structure and needs far less labeled data to reach the same accuracy. For inertial data, the most effective approach is masked reconstruction, where the model recovers deliberately hidden parts of the signal \cite{haresamudram2020masked, xu2021limu}. Time-series foundation models take this a step further, pretraining one encoder on a large collection of public series so that it can be reused, frozen, on a signal it never saw~\cite{goswami2024moment}.

\emph{Boxing biomechanics and wearable sensing.} Applying that sensing and modeling stack to boxing draws on two bodies of work: the biomechanics that describes what separates skilled punching from unskilled, and the wearable systems built for the sport so far. On the first, studies report the force, velocity, and segmental timing of punches, and how each changes with skill. Olympic-level straight punches reach impact forces of several thousand newtons and hand velocities near nine meters per second, with most of the force coming from the effective mass of the whole body rather than the arm alone \cite{walilko2005biomechanics}. A force-plate study of all six punches finds rear-leg drive critical to punch effectiveness, with the cross producing the largest ground reaction force \cite{stewart2025role}. Lower-body strength and power, in turn, predict punch impact in elite boxers \cite{loturco2016strength}. Skill shows up in the timing. More skilled boxers drive the legs and trunk before the arm, a proximal-to-distal sequence, and reach higher segment velocities than less skilled boxers \cite{dinu2020biomechanical, liu2022biomechanics}. 

On the second, the wearable systems built for boxing mostly use machine learning to recognize punches automatically. Most place inertial sensors on the wrists or upper body and classify which punch was thrown. One study compared sensor placements and supervised classifiers, reaching about $0.90$ accuracy and close to $0.98$ with the best model \cite{worsey2020evaluation}. Other work classified punch type and estimated punch range from wrist sensors during bouts \cite{manoharan2023punch} and later reduced the labeling cost of this pipeline with an active-learning scheme that held above $0.91$ accuracy while labeling only a small fraction of the data \cite{manoharan2025active}. A smaller line of work moves past punch type toward performance, using a validated hand-mounted wearable to compare experienced and non-experienced boxers on punch force, velocity, and trajectory \cite{menzel2021application}. A recent review of wearable combat-sports sensing covers technique classification, force measurement, and training-load monitoring \cite{xue2025limb}. Most of these systems measure just the wrist or upper body rather than the full kinetic chain and stop at a punch label or a force value. For an athlete trying to improve their technique, the next step is an assessment of how well each punch is thrown and feedback on what to change.

\emph{Explainable AI (xAI).} Feedback of that kind is an explanation. A model trained for label accuracy does not reveal which channels, timesteps, or body segments drove its prediction. To tell an athlete what to change, that prediction has to become a statement about the movement behind it. This is the problem xAI addresses, and its methods have grown along three lines. The first traces a prediction back to its inputs. Attribution methods identify the signals or regions that most influenced a decision \cite{selvaraju2017grad, sundararajan2017axiomatic, lundberg2017unified, ribeiro2016should, dhurandhar2018explanations}, and measures such as perturbation-based fidelity test whether those maps track the model rather than only looking convincing \cite{petsiuk2018rise, hooker2019benchmark, arrieta2020explainable, samek2021explaining, kalasampath2025literature, tjoa2020survey}. Even a faithful map, though, shows only where the model looked and leaves open what the athlete should do differently. The second addresses that gap by looking at what could have been different. Counterfactual methods describe the smallest change to the input that would flip the decision \cite{wachter2017counterfactual, mothilal2020explaining}, with quality judged on the validity, sparsity, and proximity of that change \cite{guidotti2024counterfactual}. The change is still in the raw signal, though, not in the words an athlete or coach would use. The third puts the explanation into language, using language models \cite{lewis2020retrieval, jing2018automatic}. Evaluation measures include faithfulness checks on whether the text reflects the model's basis for the decision \cite{jacovi2020towards} and qualitative human assessments of whether the wording is clear and useful \cite{van2021human}.

The three approaches suit different audiences. The attribution map speaks to the engineer or domain expert who can understand the underlying signals, though on its own it shows where the model looked without giving the transparency needed to act \cite{ghassemi2021false}. The counterfactual and the narrative serve audiences who need the assessment as a concrete change or in plain language rather than as raw signals. One form of explanation cannot serve all audiences, because an explanation is audience-relative \cite{miller2019explanation}.

\emph{Positioning.} SPAR closes the three gaps these lines of work leave open. It instruments the upper- and lower-body kinetic chain with an eight-IMU garment and a pair of pressure insoles, solves the eight orientations into anatomical joint angles through an OpenSim musculoskeletal model calibrated to each participant, automatically classifies each punch as expert or novice on those angles, and explains that prediction rather than returning a label alone (Fig.~\ref{fig:overview}). The expert/novice estimate stands in for skill: what separates the two classes is how they move, and that difference is what the explanation describes.

A skill assessment in a gym reaches three kinds of user, and each reasons at a different level. An analyst refining the system looks at the joint angles directly and wants to know which joints drove the prediction. A coach on the training floor thinks in regions of the body and needs the one link in the chain to cue. An athlete wants a plain-language description of what went wrong and how to fix it. SPAR delivers three tiers of the same prediction, one for each. Contrastive Ablation Attribution (CAA) attributes the decision to each joint and plantar zone by permuting that unit's series across punches and measuring the resulting change. Directed Counterfactual Search (DCS) keeps the single-change logic of a counterfactual but searches over the kinetic-chain layers in place of raw channels. Grounded Narrative Generation (GNG) has a language model put the outputs of the other two tiers into plain language.

Analyzing what drives the expert/novice decision reveals what separates expert technique from novice. SPAR treats an explanation of the prediction as feedback on technique. This only holds if the classifier separates the classes by how a punch was thrown. In Study~1, we collect the dataset in a boxing gym. On that dataset we evaluate the classifier with leave-one-participant-out (LOPO) cross-validation and audit the two quantitative tiers. In Study~2, six practicing boxing coaches use all three tiers and we assess whether the explanations are useful to them. To our knowledge, SPAR is the first wearable movement-assessment system to combine full-body kinetic-chain sensing, per-punch skill assessment, and a three-tier explanation framework, with data collected in an ecologically valid training setting. 

This paper provides four contributions:
\begin{itemize}
  \item \textbf{Full-body wearable system:} An eight-IMU garment and a pair of pressure insoles that instrument the kinetic chain from the feet to the arms, a musculoskeletal model calibrated to each participant's T-pose that converts the sensor orientations into anatomical joint angles, and a classifier that passes those angles through a frozen time-series foundation model and labels each punch as expert or novice.
  \item \textbf{Real-world boxing dataset:} 4{,}713 segmented punches from 17 participants (5 expert, 12 novice), recorded during shadowboxing in a boxing gym and covering all six canonical punch types.
  \item \textbf{Three-tier explanation framework:} One prediction explained for three audiences, with CAA attributing it to eight joints and six plantar zones for the analyst, DCS naming the kinetic-chain layer whose correction would most raise it for the coach, and GNG turning that into a paragraph for the athlete.
  \item \textbf{Coach evaluation:} Semi-structured interviews in which six practicing boxing coaches used all three tiers, analyzed thematically into six themes.
\end{itemize}

The rest of the paper is structured as follows. Section~\ref{sec:methods} describes the wearable system, the data-collection study, the signal processing and skill classifier, the three-tier explanation framework, and the coach-evaluation study. Section~\ref{sec:results} reports classification, xAI audit, and user-study results. Section~\ref{sec:discussion} discusses the results and limitations, and Section~\ref{sec:conclusion} concludes.

\section{Materials and Methods}
\label{sec:methods}

\subsection{Wearable System}
\label{sec:system}

We developed SPAR to instrument the full kinetic chain. The system comprises three parts: an instrumented garment, a pair of pressure insoles, and a browser-based collection application.

\begin{figure}[t]
    \centering
    \subfloat[\centering\label{fig:garment:shirt}]{\includegraphics[width=0.4\linewidth,keepaspectratio]{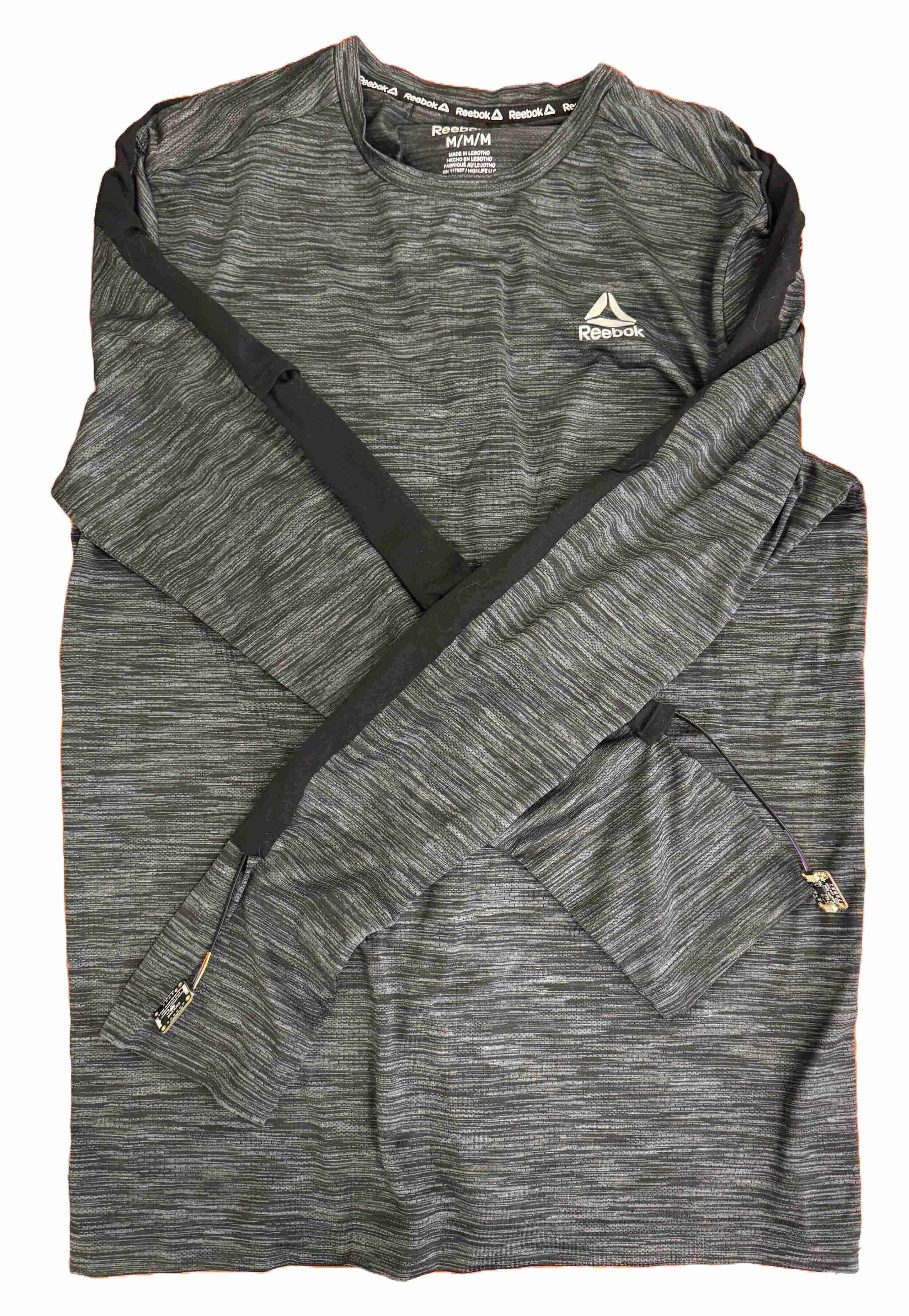}}
    \hfill
    \subfloat[\centering\label{fig:garment:pant}]{\includegraphics[width=0.4\linewidth,keepaspectratio]{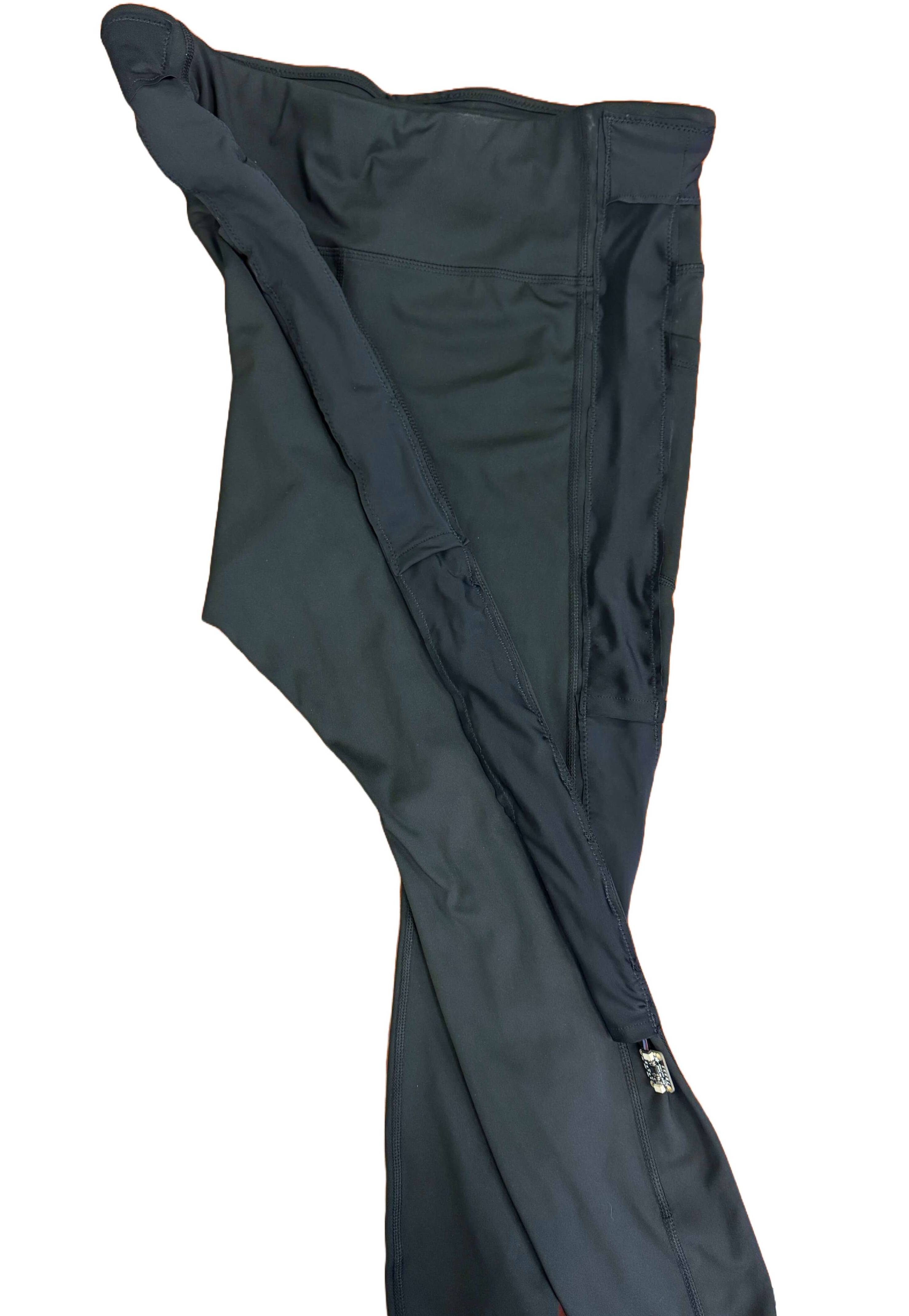}}
    \caption{Two-piece instrumented garment, with sensor placement shown on the body in Fig.~\ref{fig:overview}. (\textbf{a}) Upper-body shirt, carrying the bilateral shoulder, bilateral elbow, thorax, and sacrum IMUs. (\textbf{b}) Lower-body pants, carrying the bilateral quadriceps IMUs.\label{fig:garment}}
\end{figure}

Eight Bosch BNO055 nine-axis IMUs are placed in a two-piece garment, an upper-body shirt and lower-body pants (Fig.~\ref{fig:garment}). The placement follows the biomechanics literature on punch force and timing. A review of punch biomechanics identifies trunk rotation and lower-extremity drive as the dominant contributors to punch force \cite{lenetsky2013assessment}, so sensors are placed on the thorax, sacrum, and bilateral quadriceps. A kinematic study shows that proximal-to-distal sequencing separates elite from junior boxers \cite{dinu2020biomechanical}, so bilateral shoulder and elbow sensors observe the arms. All eight mount on bone-close landmarks (distal radius, lateral humerus, spinous processes, sacrum, distal lateral femur) to minimize soft-tissue artifact \cite{camomilla2018trends}.

A single I\textsuperscript{2}C bus connects the eight sensors to an Adafruit ESP32-C6 Feather, which is worn on the lower back, runs off a small battery, and streams their unit quaternions over a WebSocket to a laptop at a nominal 100\,Hz. The achieved rate is lower and varies between participants. The median inter-sample interval corresponds to $87$\,Hz, per-participant means run from $61$ to $89$\,Hz, and the 99th-percentile interval of $21$\,ms indicates occasional dropped samples on the shared bus. Each IMU runs in six-degree-of-freedom fusion mode with the magnetometer disabled, since indoor training spaces hold enough magnetic interference to corrupt the heading estimate. The garment does not record raw acceleration. The extra three channels per sensor would raise the payload by three quarters on a link that already drops samples, and a slower rate would cost the sequencing the system is built to measure.

Inside the participant's footwear, a pair of Novel Loadsol-3 pressure insoles streams at 100\,Hz to a paired iPhone \cite{burns2019validation}. Each foot reports three plantar pressure zones (heel, midfoot, forefoot) and their total force. Our system uses the six zone forces only. A wall-clock anchor written at session start aligns the IMU and insole streams.

\begin{figure}[t]
  \centering
  \includegraphics[width=\textwidth,keepaspectratio]{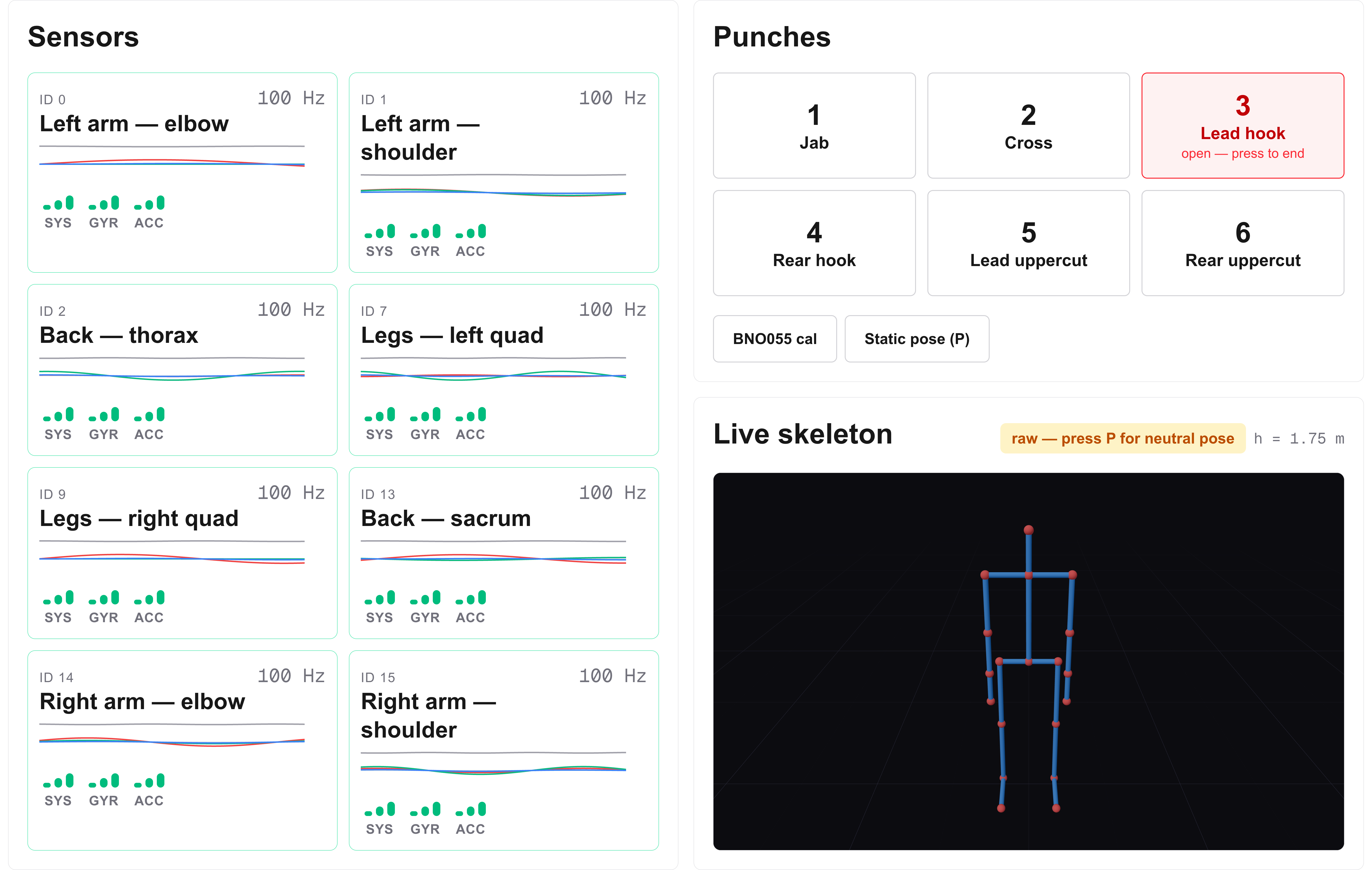}
  \caption{Browser-based data-collection and review application: live 3-D skeleton viewer, WebSocket status, per-sensor calibration display, and punch-type labeling controls.}
  \label{fig:webapp}
\end{figure}

Data collection runs through a browser-based application that records the IMU stream, exposes a labeling interface for the six canonical punch types (jab, cross, lead/rear hook, lead/rear uppercut), and renders a live three-dimensional skeleton from the sensor orientations for checking data quality during the session (Fig.~\ref{fig:webapp}). The insoles record in parallel through Novel's loadapp on the iPhone.

\subsection{Study 1: Data Collection}
\label{sec:study1}

\subsubsection{Participants and IRB}
\label{sec:study1:participants}

The Institutional Review Board of Vanderbilt University Medical Center approved this work under protocol 241478. Eligibility required age $\geq$\,18, clearance for moderate physical activity, and no upper-limb injury within six months of the session. We stratified participants into experts (three or more years of competitive boxing experience) and novices (one year or less of training). The final cohort comprised 5 experts (E1--E5) and 12 novices (N1--N12). Eleven were male and six were female.

\subsubsection{Session Protocol}
\label{sec:study1:session}

After consent, we measured the participant's height, fitted the garment, calibrated the sensors, and ran a single recording block.

Calibration ran in two stages. First, we swept the garment through a figure-8 motion until every BNO055 reported full fusion calibration. Second, the participant held an anatomical T-pose for five seconds, and a trimmed eigenvector mean~\cite{markley2007averaging} of each sensor's window gives a per-sensor rest quaternion $q^\mathrm{rest}_s$. That reference registers each sensor to its body segment at model calibration (Section~\ref{sec:model:features}) and fixes each sensor's heading, which is arbitrary at power-on with the magnetometer disabled. Drift after that point is left uncorrected (Section~\ref{sec:discussion:limits}).

Every participant boxed in the orthodox stance and threw the six punches in a fixed order, jab, cross, lead hook, rear hook, lead uppercut, and rear uppercut, with as many repetitions of each as they wanted and rest between types. The first author, with seven years of boxing experience, marked each punch with a keypress that recorded its type and, through the press-to-release interval, its start and end.

\subsubsection{Resulting Dataset}
\label{sec:study1:dataset}

We collected 4{,}713 individually segmented punches from the 17 participants, spanning all six canonical punch types, after discarding sub-100\,ms segments as noise. Table~\ref{tab:dataset} reports the per-participant breakdown by identifier, sex, height, per-punch-type counts, and mean per-punch duration.

\input{tables/table_dataset.tex}

\subsection{Signal Processing and Skill Classifier}
\label{sec:model}

SPAR's classifier takes one punch's sensor quaternions and insole channels and labels it expert or novice. It first solves the orientations into anatomical joint angles, then encodes each joint-angle and plantar-force series with a frozen pretrained time-series model, and classifies the encoded series with a small transformer trained on the cohort. The series stay separate through the encoder, so CAA can attribute the prediction to individual joints and DCS can substitute groups of them.

\subsubsection{Preprocessing}
\label{sec:model:pre}

Each punch is resampled to a uniform 100\,Hz grid, which equalizes the variable sample spacing across participants (Section~\ref{sec:system}) without adding information the sensors did not capture. Linear interpolation would pull quaternions off the unit sphere, so orientations are resampled with SciPy's RotationSpline, a minimum-jerk spline fit in the tangent space of SO(3), which keeps every resampled value a valid rotation. Insole channels are resampled by piecewise-linear interpolation and z-scored per participant, which removes the body-weight, foot-size, and shoe-type baselines.

We do not rescale punches to a common length. Duration could itself be a skill signal, since the fastest participant throws a cross in 492\,ms against a cohort mean of 1041\,ms, so every series is encoded over a punch's own samples and duration is passed to the classifier as its own token.

\subsubsection{Representation}
\label{sec:model:features}

The classifier's input is 28 time series per punch, the 22 joint angles and the six plantar zone forces, each encoded on its own (Fig.~\ref{fig:ml-pipeline}). The joint angles are described first, then the encoding.

\emph{From orientation to joint angles.} Each participant's T-pose registers every sensor to its body segment through OpenSim's IMU calibration~\cite{delp2007opensim, al2022opensense}. The musculoskeletal model is the Rajagopal full-body model~\cite{rajagopal2016full}, posed in the same T-pose, with its upper-body coordinates unlocked and its joint ranges widened for boxing. Inverse kinematics then fits the model to the eight orientations frame by frame and returns each segment's rotation relative to its parent along anatomical axes. The pelvis and lumbar spine each have three (tilt, list and rotation for the pelvis; extension, bending and rotation for the lumbar spine), each hip and shoulder has three (flexion, adduction and rotation), and each elbow has two (flexion and forearm rotation), for 22 coordinates from eight sensors. The knees are left out, since no sensor constrains them. A hip flexion angle means the same thing on every participant regardless of how the thigh sensor was strapped.

\emph{Sensors that stop reporting.} The garment is programmed such that a sensor that loses its connection repeats its last value. To account for this, we detect the repeated value by near-identical consecutive samples over most of the punch and replace that sensor's stream for the punch with its T-pose reading, so the solver holds the segment at its calibration pose rather than inventing motion for it.

\emph{Encoding.} Each of the 28 series is encoded separately by MOMENT~\cite{goswami2024moment}, a transformer pretrained by masked reconstruction on a large public collection of time series, with its weights frozen. MOMENT operates on a 512-sample window in patches of eight samples, and at 100\,Hz a punch covers only about twelve of them. The lags between joints are tens of milliseconds, which at that rate fall inside a single patch where the encoder cannot resolve them. Each series is therefore upsampled by a factor of three and zero-padded to the window, with an input mask that hides the padding from attention. Three is the largest factor at which the 95th-percentile punch of 170 samples still fits the window, and it is the same for every punch, so durations and the timing between joints are preserved. MOMENT also standardizes each series to zero mean and unit variance before encoding it, which discards the angle a joint was held at and the range it moved through and keeps only the shape of the trajectory. Posture and range of motion separate the classes, so we disable that step and instead shift and scale each channel by one constant pair fitted over the corpus. The encoder returns one 512-dimensional embedding per patch, and the patches covering the punch are averaged into eight equal phase bins, so a punch becomes $28 \times 8 = 224$ tokens.

Foundation models for inertial sensing appeared while we developed SPAR~\cite{zhang2024unimts, miao2026wonderwall, xu2026inertia}. Their input is accelerometer and, in some cases, gyroscope signals, whereas we chose to represent each punch as the anatomical joint angles a musculoskeletal model reports, following the biomechanics literature. A general time-series encoder takes the joint angles as they are, whereas an inertial one would need them converted back into accelerations and angular velocities. These models are also typically pretrained at $20$\,Hz~\cite{zhang2024unimts, xu2026inertia}, and downsampling our data to that rate is restrictive for boxing. A sample period of $50$\,ms cannot capture the lag between successive layers' peak joint speeds, which is under $30$\,ms in this cohort and differs between the two classes by about $20$\,ms.

\subsubsection{Classifier}
\label{sec:model:arch}

\begin{figure}[t]
  \centering
  \includegraphics[width=\textwidth]{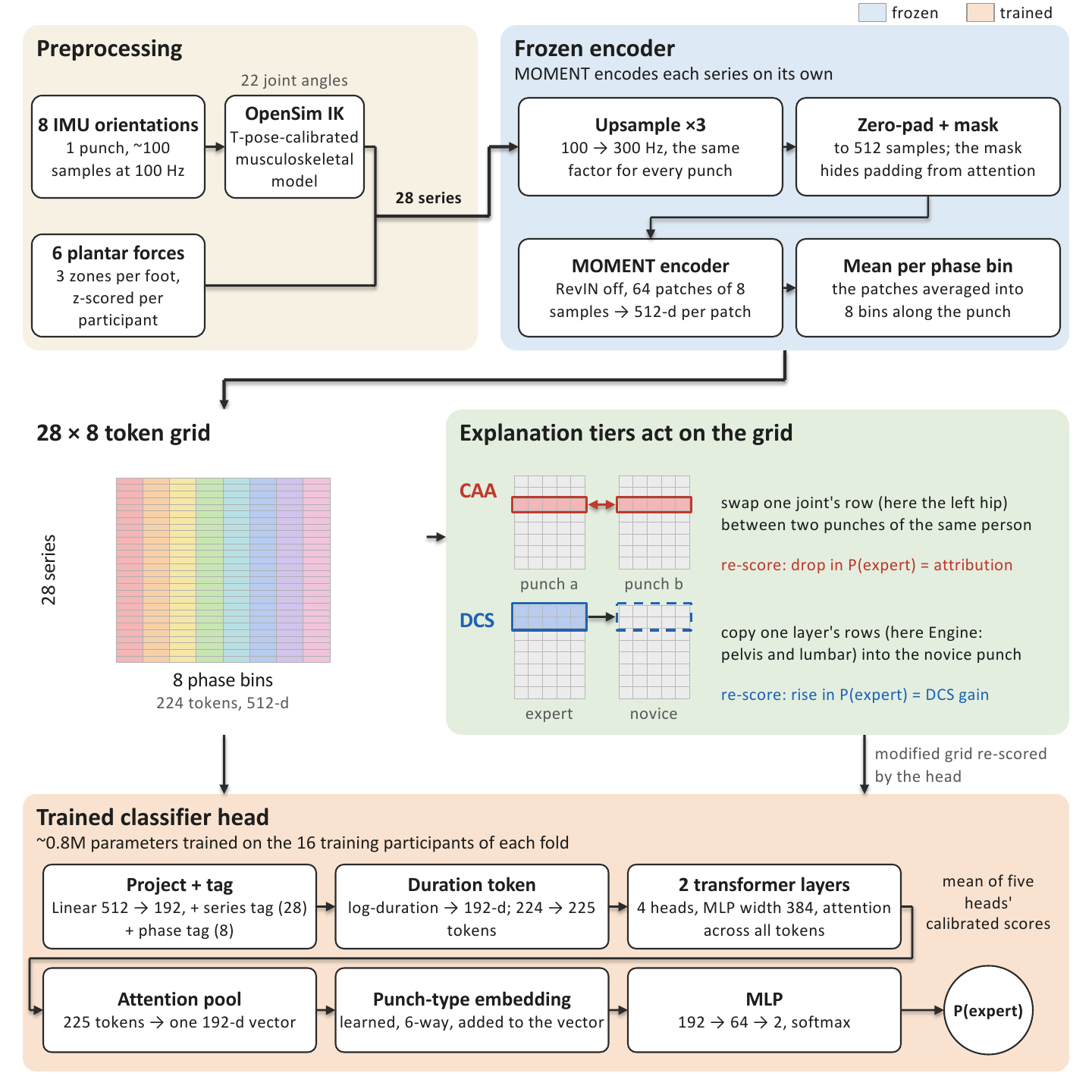}
  \caption{The skill classifier. Top: the eight orientations are solved into 22 joint angles and combined with the six plantar zones to get 28 series. Each series is encoded on its own by the frozen MOMENT encoder and averaged into eight phase bins, giving a 28~$\times$~8 grid of tokens. Bottom: the trained head, averaged over five seeds. Right: CAA swaps one joint's rows between the participant's punches and DCS copies one layer's rows from the matched expert punch, and the modified grid is re-scored by the head.}
  \label{fig:ml-pipeline}
\end{figure}

The classifier is a small transformer that maps the 224 tokens to a novice or expert decision (Fig.~\ref{fig:ml-pipeline}, bottom). We project each token to 192 dimensions and add two learned tags, one for the series and one for the phase bin, so the model knows which joint and which part of the punch each token came from. We also add a token for the punch's log-duration, since the phase bins remove duration. Two transformer layers with four heads attend across the tokens. An attention pool then collapses them to one vector, we add a learned embedding of the punch type, and a two-layer perceptron ends in two output nodes, one for novice and one for expert. A softmax over the two gives the class probabilities, and the expert probability is the score we use throughout. The head has ~0.8M trainable parameters. The encoder's 35M stay frozen.

We train the head in each fold with AdamW (learning rate $3\times10^{-4}$, weight decay $10^{-4}$) for 20 epochs on class-balanced minibatches of 32 with label smoothing of 0.1. During training we zero all of a series' tokens with probability 0.2, so the head cannot rely on one joint. With 16 training participants, different random seeds reach different solutions, so we train five heads and average their calibrated scores.

\subsubsection{Training and Evaluation Protocol}
\label{sec:model:train}

The skill label is a property of the participant, so a held-out fold is single-class, and its punches are compared against opposite-class punches pooled from the other folds. We report AUC rather than accuracy, since accuracy needs a threshold the model cannot calibrate for an unseen person.

\emph{Score calibration.} Each fold trains its own head and heads differ in how confident they are, so we convert each held-out score to a percentile before pooling. The percentile is the fraction of the fold's training punches that the punch scores above. The reference scores must come from that fold's training participants but not from the fold's own head, which is overconfident on punches it has seen. So within each fold we score the training punches with a nested cross-validation over the training participants, in which every training punch is scored by a head that never saw its participant. With $c(x)$ the percentile of punch $x$ and $\mathcal{E}_p$ and $\mathcal{N}_p$ the expert and novice punches compared for participant $p$ (their own punches against the opposite-class punches from the other folds),
\begin{equation}
  \mathrm{AUC}_p = \frac{1}{|\mathcal{E}_p|\,|\mathcal{N}_p|} \sum_{e \in \mathcal{E}_p} \sum_{n \in \mathcal{N}_p} \mathbb{1}\!\left[c(e) > c(n)\right],
  \label{eq:auc}
\end{equation}
and the main result is the mean of $\mathrm{AUC}_p$ over the 17 participants.

\emph{Uncertainty.} The label is a participant-level property, so the bootstrap is run over the 17 participants and a 95\% interval is reported throughout. Variation across seeds is reported separately.

\subsection{Three-Tier Explanation Framework}
\label{sec:xai}

The expert/novice prediction stands in for the difference in how the two classes move, so explaining what drives it exposes that difference. SPAR does so with three tiers, each for a different audience: CAA attributes the prediction to individual joints for the analyst, DCS searches the kinetic-chain layers for the coach, and GNG writes the two up in plain language for the athlete.

CAA works at the level of individual joints and plantar zones. DCS and GNG group them into three layers along the proximal-to-distal kinetic chain documented in boxing biomechanics, where force begins in the legs, passes through the trunk, and the arm delivers it~\cite{lenetsky2013assessment, dinu2020biomechanical}. The grouping and the layer names are ours and follow how coaches describe the chain. Foundation covers the two hips and the insoles, which capture leg drive, stance, and ground-contact sequencing. Engine covers the pelvis and lumbar spine, which capture pelvic rotation, trunk rotation, and inter-segment timing. Delivery covers the two shoulders and two elbows, which capture arm extension, rear-arm sequencing, and retraction.

\subsubsection{Contrastive Ablation Attribution}
\label{sec:xai:caa}

CAA ranks the eight joints and the six plantar zones by how much the prediction changes when each is removed.

We remove one joint at a time by shuffling its tokens across the held-out participant's punches. Each punch is randomly paired with another of that participant's punches and takes that punch's tokens for joint $j$, keeping its own tokens for every other joint and its own duration and punch type. Joint $j$'s tokens still come from a real punch by the same person but no longer match the rest of the punch, so whatever the classifier drew from that joint is removed. The classifier is then re-applied to every punch. With $f(x)$ the classifier's expert probability and $\Pi_j$ the shuffle,
\begin{equation}
  \phi_j = \mathbb{E}\!\left[f(x)\right] - \mathbb{E}\!\left[f(\Pi_j x)\right],
  \label{eq:caa}
\end{equation}
with both expectations over the held-out participant's punches. We repeat this with five random pairings and average. Five are enough that the standard error of $\phi_j$ from the shuffling is a few thousandths, below the differences between the leading joints. The procedure repeats for each of the eight joints and, in the same way, for each of the six plantar zones. We report $|\phi_j|$ per fold, averaged over the folds of each class, since a joint can move the prediction in either direction.

\subsubsection{Directed Counterfactual Search}
\label{sec:xai:dcs}

DCS identifies the one layer whose replacement with an expert's most raises the expert probability. It runs on the novice held-out folds.

Replacement is done one novice punch at a time. First, an expert punch is chosen from the fold's training set to supply the replacement. The novice punch is compared to every training expert punch on the tokens of the series outside the layer, standardized over the training set, and the expert punch at the smallest Euclidean distance is taken. Second, the layer's tokens are copied from the expert punch into the novice punch. A layer consists of its joints' series, and for Foundation the six plantar zones as well. Third, the classifier is re-applied.

With $x$ the novice punch and $x^{(k)}$ the same punch with layer $k$'s tokens replaced by those of its matched expert,
\begin{equation}
  \Delta p_k = f\!\left(x^{(k)}\right) - f(x).
  \label{eq:dcs}
\end{equation}
$\Delta p_k$ is averaged over the participant's punches, and the layer with the largest rise is their primary gap~\cite{wachter2017counterfactual, mothilal2020explaining, verma2024counterfactual}.

\subsubsection{Grounded Narrative Generation}
\label{sec:xai:gng}

GNG puts the two quantitative tiers into plain language for the athlete. The payload for one punch has four parts. It gives the punch type and the classifier's expert probability. It lists the CAA attribution of each joint and plantar zone for that participant, in ranked order. And it lists the DCS rise in expert probability for each layer, in ranked order, with the largest named as the primary gap.

A language model (Claude Sonnet 4.5) receives the payload together with a short prompt. The prompt defines each layer in body terms (Foundation is the hips and feet, Engine the pelvis and trunk, Delivery the shoulders and elbows), asks for one short paragraph in the second person that names the region to work on first and the joints the assessment rested on, and forbids any claim, cue, or drill that is not in the payload. The model answers in two parts, a one-line restatement of the primary gap and the three most attributed joints, then the paragraph. A check compares the restatement with the payload, confirms that no other region is named as the first thing to fix, and regenerates on a mismatch. Grounding rests on the payload containing only the outputs of the other two tiers, on the prompt confining the model to it, and on that check. The payload could carry much more, such as a library of common faults with matching drills or the athlete's history, and the same check would apply to each addition. We keep it simple here, since the aim of this paper is to demonstrate the three-tier framework rather than the richest narrative the third tier could produce.

\subsection{Study 2: Coach Evaluation}
\label{sec:coach}

\begin{figure}[t]
  \centering
  \includegraphics[width=\textwidth,keepaspectratio]{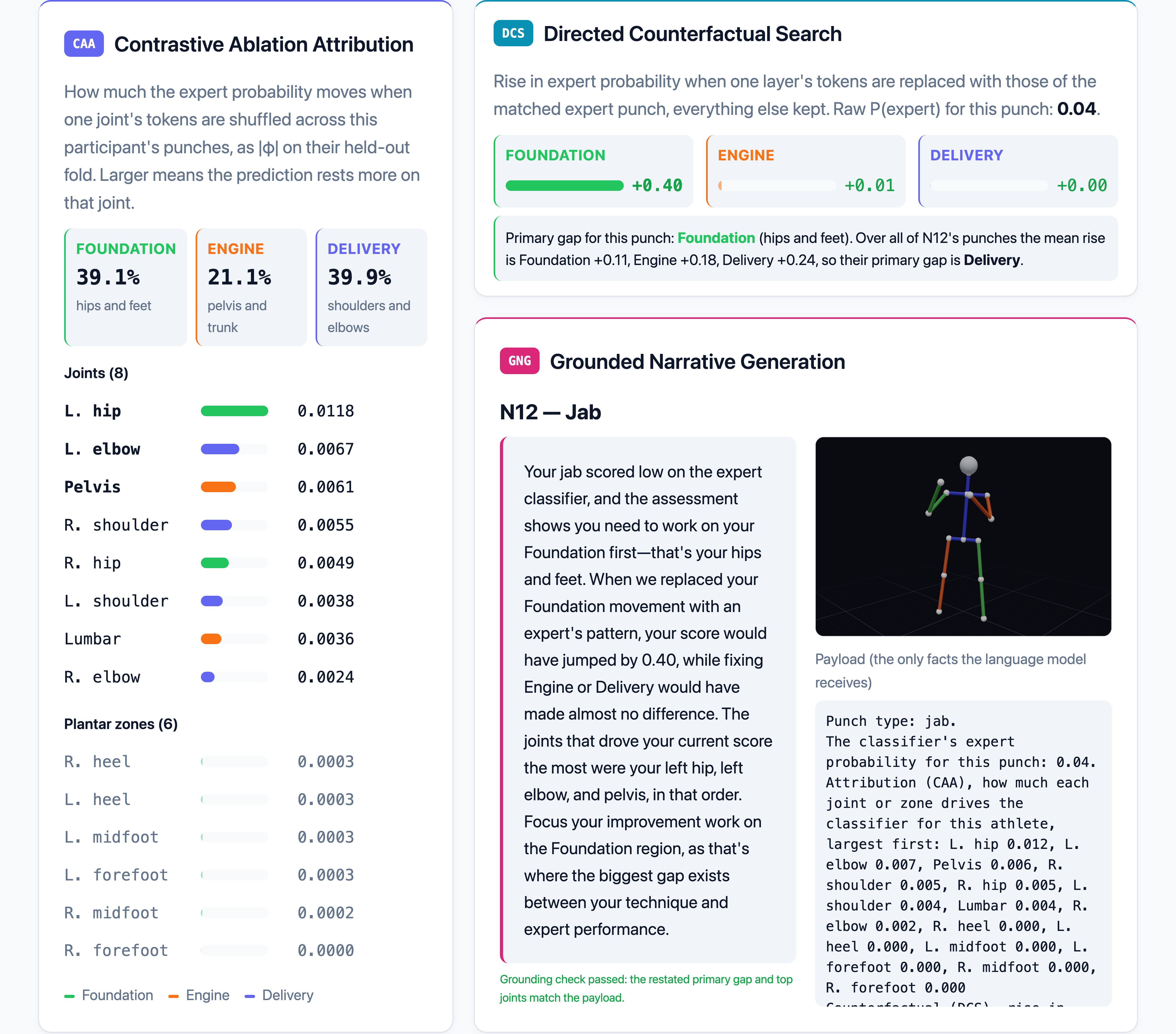}
  \caption{Coach-evaluation viewer used in Study~2. Coaches replay a recorded punch in the 3-D skeleton view and inspect the three system outputs together, the CAA channel attribution, the DCS counterfactual, and the GNG narrative.}
  \label{fig:coach_viewer}
\end{figure}

Study~2 examines how six boxing coaches use SPAR through a thematic analysis of semi-structured interviews~\cite{braun2006thematic}. The coaches were recruited from boxing and combat-sport gyms in the Nashville metro area and have 4 to 20 years of experience across three corporate gym formats. Four run group classes in large gyms (C1, C2, C4, C5), with C1 and C4 in youth programs and C2 in adult classes. Two work in a technique-analysis role with competitive athletes (C3, C6). Four carry trainer-certification duties (C1, C3, C4, C6). Each coach gave informed consent under the same protocol and sat for one in-person semi-structured interview of 25 to 30 minutes, working in a browser-based viewer (Fig.~\ref{fig:coach_viewer}) that showed recorded punches next to the three system outputs. We asked what each explanation was saying, whether the coach would have flagged the same issue, and how they would use the system in their own gym.

We transcribed the recordings and followed the six phases of reflexive thematic analysis~\cite{braun2006thematic}, which is carried out by one analyst. The coding produced 36 initial codes, grouped into six themes and checked against the transcripts over repeated passes.

\section{Results}
\label{sec:results}

Results cover the skill classifier (Section~\ref{sec:results:classifier}), the audit of the two quantitative explanation tiers (Section~\ref{sec:results:xai}), and the thematic analysis of the six coach interviews (Section~\ref{sec:results:userstudy}).

\subsection{Skill Classifier}
\label{sec:results:classifier}

The classifier reaches a mean-over-participants LOPO AUC of $0.842 \pm 0.097$ across the five seeds (bootstrap 95\% CI $[0.769, 0.907]$ over participants). Table~\ref{tab:perfold} gives the per-participant breakdown.

\input{tables/table_per_fold.tex}

Thirteen of the seventeen exceed $0.70$ and seven exceed $0.90$; the three lowest are E1 at $0.503$, N1 at $0.595$ and N10 at $0.668$. Expert and novice folds reach about the same AUC, $0.869$ and $0.830$ on average.

\subsubsection{Ablations}
To assess the contribution of the encoder-side choices, we removed each one in turn and re-ran the LOPO evaluation under the reported protocol. Keeping MOMENT's per-series normalization reduces the AUC to $0.523 \pm 0.048$, close to chance, since the embedding then carries only the shape of each trajectory and not the posture or range of motion that separate the classes. Encoding the punches at their native rate rather than upsampled by three reduces it to $0.724 \pm 0.050$. Replacing the encoder with the mean and standard deviation of each series in each phase bin, fed to the same head, reduces it to $0.736 \pm 0.021$.

\subsection{Explanation-Quality Audit}
\label{sec:results:xai}

\begin{figure}[t]
  \centering
  \includegraphics[width=\columnwidth]{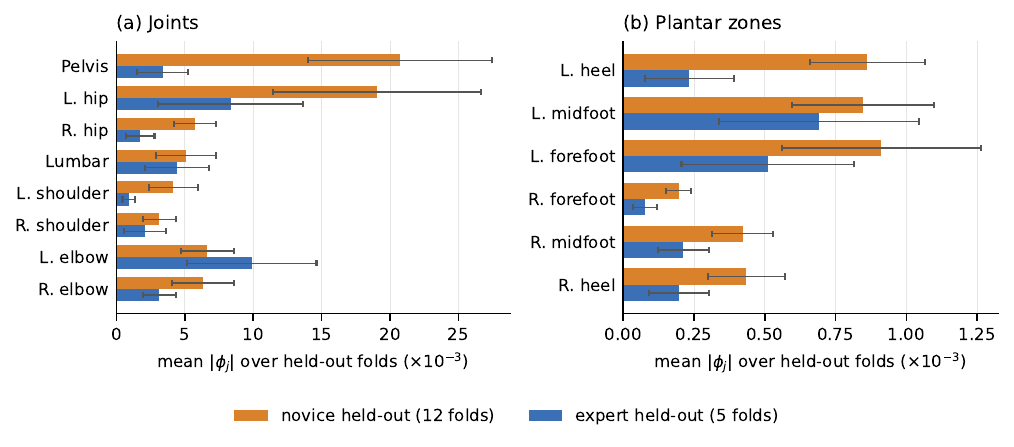}
  \caption{Mean CAA attribution per unit under LOPO, split by the held-out participant's class. (a) The eight joints. (b) The six plantar zone forces.}
  \label{fig:sensor-attribution}
\end{figure}

We audit the two quantitative tiers on the LOPO classifier. Table~\ref{tab:xai} reports the CAA attribution of each joint and plantar zone and the DCS layer substitutions. Fig.~\ref{fig:sensor-attribution} shows the attribution split by held-out class.

\input{tables/table_xai.tex}

\subsubsection{Attribution}
On novice held-out folds the pelvis and the left hip carry the largest attributions, $0.0207$ and $0.0191$, about three times the next group of joints. The left elbow, right elbow, right hip and lumbar spine follow between $0.0067$ and $0.0051$, and the shoulders sit at or below $0.0042$. On expert held-out folds every attribution is smaller and the left elbow leads at $0.0099$, followed by the left hip at $0.0083$ (Table~\ref{tab:xai}, Fig.~\ref{fig:sensor-attribution}). The spread across folds is wide relative to the gaps between neighboring joints, so those orderings are not resolved on either class. The plantar zones sit an order of magnitude below the joints, the highest at $0.0009$.

\subsubsection{Counterfactual}
Averaged over the novice folds, replacing any one layer with the matched expert's raises the expert probability by about the same amount, $+0.274$ for Foundation, $+0.272$ for Delivery and $+0.271$ for Engine. The primary gap differs by participant (Table~\ref{tab:dcs}). It is Delivery for five of the twelve novices, Engine for four and Foundation for three, and for a given novice it is often large. Foundation raises N9 by $+0.624$, Engine raises N7 by $+0.542$ and Delivery raises N8 by $+0.465$.

\input{tables/table_dcs.tex}

\subsubsection{Agreement}
In the aggregate the two tiers point at the same region. The pelvis and hips carry the largest attributions on novice folds, and the two layers that contain them, Foundation and Engine, are the primary gap for seven of the twelve novices. Per participant the agreement is weaker. Table~\ref{tab:dcs} lists each novice's most attributed unit beside their primary gap. The attribution ranking is much the same for every novice, with the pelvis or a hip on top for nine of them, whereas the primary gap moves between all three layers, so the unit CAA ranks first is often outside the layer DCS names. That difference is expected rather than a problem, because the two tiers answer different questions. CAA reports which series the prediction rests on as thrown, and that is the trunk and hips for almost everyone, since those are the series that separate the classes. DCS reports which region an expert's version would change most for one person, and that depends on where that person falls short. A joint can be what the classifier looks at without being what a given novice most needs to fix, which is why the two tiers are reported separately.

\subsection{User Study}
\label{sec:results:userstudy}

The thematic analysis of the six interviews produced 36 initial codes and six themes.

\subsubsection{Theme 1: DCS lets a coach reach more athletes}
All six coaches said DCS was the tier they would use most, because it lets them reach more athletes. In a class of 15 to 30 the limit is not what a coach knows but how many people they can watch closely in one session. DCS names the single highest-leverage fix for each athlete, so the coach can go straight to the cue. C1, who runs a youth program, said, \emph{``The limiting factor in youth coaching is how many kids I can observe in single class. This helps me quickly see what is going on.''}

\subsubsection{Theme 2: What fits depends on the coach}
Coaches mapped the three tiers to audiences much as SPAR intends, but they split them by coaching role rather than by athlete skill. The four group-class coaches (C1, C2, C4, C5) said CAA had no place in class (\emph{``I don't look at movement that way,''} C4). The two who work in a technique-analysis role (C3, C6) would use CAA the other way round, to check their own understanding of a punch after the fact (\emph{``A punch happens so quickly. This lets me look at it more objectively.''} C6). Every coach said they would use DCS in class and GNG as a take-home tool. DCS fit the floor because it gives one cue per athlete that a coach can act on between rounds without stopping to interpret a display. GNG fit the time after class because it is written text the athlete can review with their own understanding later.

\subsubsection{Theme 3: One voice does not fit every audience}
The slightly formal voice of the GNG paragraph split the coaches. Four (C1, C2, C4, C6) liked it for take-home material (\emph{``People at nice gyms want the experience to feel premium. The text being slightly formal looks good,''} C2). The other two (C3, C5) felt the same voice was wrong for talking to an athlete face to face and wanted to be able to change it, one register for a written summary and a blunter one for the floor (\emph{``different voice for different people,''} C3).

\subsubsection{Theme 4: SPAR can train new coaches}
The four coaches with trainer-certification duties (C1, C3, C4, C6) saw SPAR as a potential tool for developing coaches, separate from its use with athletes. New trainers often cannot name what they are seeing. The narrative gives them the words for it, and once they have seen the pattern enough times they no longer need the system (\emph{``With the system telling them that the issue is in the engine layer, work on hip rotation, they start to develop a vocabulary.''} C3).

\subsubsection{Theme 5: Families as part of youth coaching}
Both youth-program coaches (C1, C4) raised a use we had not planned for, the GNG paragraph as a way to keep parents involved. Both linked family involvement to kids staying in the program (\emph{``Kids stay in programs when their families get invested. If the only person who knows what the kid is working on is me, the kid doesn't get reinforcement at home.''} C4). Both were also clear that children should not see CAA displays of their own bodies (\emph{``Kids should be feeling the punch, learning to trust their bodies, not analyzing themselves,''} C1), so in a youth setting CAA is used just by the coach and the parent.

\subsubsection{Theme 6: Traceability earns trust}
Two coaches (C5, C6) said they trusted the feedback because they could see where it came from. Each assessment names the joints the prediction rested on and the region whose correction most raised it, and the narrative is written from those two facts alone, so any cue can be traced back to what produced it. C5 put it in terms of the narrative's inputs (\emph{``The summary text can be trusted more since I know what's being fed into it.''}). C6 contrasted it with AI products that vendors had pitched before (\emph{``We trust it more when we can see what the system is seeing.''}).

\subsubsection{Other feedback}
Outside the six themes, the coaches pointed out gaps that stop the system from being deployable, and most are engineering problems rather than model problems. Setup takes about 5 minutes per person now, and the coaches wanted it under a minute. For the outputs, they wanted CAA display to show direction, the GNG paragraph to have a send-to-member export, and an adjustable voice. They also wanted the system on a phone or tablet they could use while coaching rather than on a laptop, and a finer set of cues for advanced and competitive members. These requests came mostly from the two coaches who work with competitive members (C3, C6).

\section{Discussion}
\label{sec:discussion}

\subsection{A Three-Tier Explanation Framework for Multiple Audiences}
\label{sec:discussion:pipeline}

SPAR explains one skill prediction at three tiers for three audiences. We designed CAA, DCS, and GNG for those tiers, and each is defined by what it gives its audience. Any method that produces the same output could take its place. Shapley values could be used for CAA, a generative counterfactual for DCS, or a different language model for GNG. What we argue in this paper is that the structure of the explanation matters. In our case, a skill prediction is presented differently to the different people who will act on it at the level each of them reasons at.

The idea that an explanation depends on who receives it is not new. The main xAI taxonomy makes the audience part of the definition of an explanation~\cite{arrieta2020explainable}, and the social-science account holds that people expect explanations to be selected and contrastive rather than complete~\cite{miller2019explanation}. However, a system that acts on this idea is rare. Most xAI work in medicine and wearable sensing produces a single saliency map and leaves the reader to interpret it~\cite{tjoa2020survey, kalasampath2025literature}, and clinical work has argued that such maps are the wrong output for a practitioner~\cite{ghassemi2021false}.

Study~2 supports the same idea from the practitioner's side. Together with the literature above it suggests that one explanation is unlikely to serve every person who has to act on the same model output. Giving each audience its own tier that traces back to the model is a reasonable design requirement for such systems.

\subsection{A Wider Spectrum of Explanations}
\label{sec:discussion:spectrum}

The three audiences in the previous section were fixed in advance. During Study~2 we learned that there are more audiences and more relationships between them than the three we designed for. The coaches split the tiers by their own role rather than by athlete skill, two of them wanted the narrative in one voice for a written summary and another for the floor, and the youth coaches wanted the paragraph to reach parents while the attribution stayed away from the child. A framework with three fixed tiers cannot follow this, since the audiences are not fixed and new ones will appear in every gym the system enters. A more general design would let the audience be set at use rather than at design time.

GNG is a practical place for this spectrum to be developed from. The narrative model currently sees the punch only through a short text payload, a primary gap and three joint names, so everything it can say has to pass through that bottleneck. This does not have to be the case. The encodings of the punch that the classifier uses could be passed to the language model directly through a learned projection into its embedding space, which is how vision-language models take in image encoder tokens~\cite{liu2023survey} and how time-series work maps series into a language model's token space~\cite{jin2024time}. The language model would then have the movement itself to reason over rather than a summary of it. It could answer a coach's follow-up question, compare two of the athlete's punches, or describe a trend over sessions without a separate tier for each. The cost is the grounding guarantee that the coaches valued (Theme~6). With the encodings in the prompt there is no payload to check the narrative against, so a model of this kind would need to state the reasoning it relied on alongside the narrative. Reinforcement learning offers ways to train a model to reason in this way, such as group relative policy optimization, which has been used to elicit explicit reasoning in language models~\cite{guo2025deepseek}, and we leave that as an open question for future work.

\subsection{An Expert Outside the Training Distribution}
\label{sec:discussion:e1}

The explanation tiers are only as good as the classifier they explain. In our cohort, one expert performed differently from the other four. E1 has the lowest fold AUC at $0.503$. Averaged over its punches, E1's expert probability is $0.41$, the same as the average novice, whereas the other four experts average $0.73$ to $0.99$. With E1 sitting among the novices, small changes between seeds move its punches above or below theirs, which is the $\pm 0.29$ in Table~\ref{tab:perfold}.

A score this low means the classifier found little in E1's punches that it had learned to associate with experts, and we attribute this to the experts it learned from. E1's fold trains on the other four experts, and those four are more similar to each other than any of them is to E1. To quantify this, we computed the cosine similarity between each participant's mean token and the mean token of the other experts. E1 scores $0.80$ on this measure, the other four experts score between $0.91$ and $0.95$, and eleven of the twelve novices score above $0.80$ against the same four experts. E1's three nearest participants by the same measure are all novices, whereas every other expert's nearest participant is an expert. The classifier learned what those four experts look like, and E1 does not look like them. This is a consequence of the cohort size and more experts would address it.

\subsection{Beyond Boxing}
\label{sec:discussion:beyond}

The work presented here revolves around boxing, but the explanation framework is more general. It applies wherever a prediction has to be explained to several kinds of people. Clinical and rehabilitation settings are a growing example, since the same prediction there may have to be explained differently to each of the people who rely on it.

In a gait clinic, for example, a system might flag a patient's gait as a fall risk. That prediction could be used by many different people. A biomechanist wants to know which joints the model relied on, a physical therapist wants to know what to work on in the next session, and the patient needs to hear it in plain language. A family member, a referring physician, or an insurer may each need something different again. The same holds for rehabilitation exercises, occupational movement, or any other setting where the people who act on a prediction need to trust it.

This is also how the framework relates to the wider xAI landscape. The field is usually organized by method, with attribution methods such as SHAP and Grad-CAM~\cite{lundberg2017unified, selvaraju2017grad}, counterfactual methods~\cite{wachter2017counterfactual, guidotti2024counterfactual}, and natural-language explanations forming separate lines of work that are compared on faithfulness to the model~\cite{samek2021explaining, jacovi2020towards}. SPAR does not add a method to any of these lines. It organizes them by the person they serve rather than by the technique. In this paper that meant one method per tier, and Section~\ref{sec:discussion:spectrum} describes a version in which the narrative tier covers more of the range on its own. In either case the existing xAI literature supplies the parts, and the framework is a way of arranging those parts around the people who will use them.

\subsection{Limitations}
\label{sec:discussion:limits}

Several limitations qualify these findings.

\subsubsection{Cohort size}
The 17 participants put us at the low end of the IMU literature for cross-subject work, and the bootstrap intervals over participants span roughly $\pm 0.07$. With five experts, each LOPO fold learns what an expert looks like from four people, and Section~\ref{sec:discussion:e1} shows what happens when one of the five does not resemble the others.

\subsubsection{Binary skill label}
Experts had three or more years of competitive boxing and novices had one year or less. Skill is a continuum, and it keeps changing from three years of experience to more, so two classes cannot place a boxer between the groups or separate two boxers within one.

\subsubsection{Punch-level labels}
Every punch has the label of the person who threw it, so an expert's poor punch is labeled expert and a novice's best punch is labeled novice. Labeling individual punches is difficult, since a punch has no outcome measure in our recordings and coaches may not agree on it.

\subsubsection{Stance}
Every participant boxed orthodox, so the lead side is the left side for the whole cohort. A southpaw would reverse this, and we have not tested whether the classifier transfers across stances.

\section{Conclusions}
\label{sec:conclusion}

Wearable boxing systems today mostly report which punch was thrown, and the lab studies that can describe technique do not follow the athlete out of the lab. SPAR addresses that gap. It instruments the full kinetic chain with eight IMUs and a pair of pressure insoles, converts the sensor orientations into anatomical joint angles through a musculoskeletal model calibrated to each participant, and classifies each punch as expert or novice with a small transformer over a frozen time-series encoder. We collected 4{,}713 punches of six types from 17 participants, five experts with three or more years of competitive boxing and 12 novices with one year or less, and the classifier reaches a LOPO AUC of $0.842$ on that data.

The prediction on its own is not useful to a coach. CAA reports which joints the prediction rested on for the analyst, DCS identifies the kinetic-chain region to correct for the coach, and GNG writes both into a paragraph for the athlete that says nothing the first two tiers did not produce. The system was validated with six practicing coaches, who said they would use the counterfactual in class settings and the narrative as take-home material and who found two audiences we had not designed for. This evidence shows that explanations need to be formatted and presented differently for different audiences.

The work presented here is not tied to boxing or to wearable sensing. Any system that conveys a prediction to different kinds of people faces the same problem, and each of those people needs an explanation they can act on.

\section{Funding}
This work was supported by the National Science Foundation (NSF) grant 2124002.

%Bibliography
\bibliographystyle{unsrt}  
\bibliography{references}

\end{document}

%% file: tables/table_dataset.tex
\begin{table}[t]
  \caption{Per-participant breakdown of the Study-1 dataset: identifier, self-reported sex (M/F), height in meters, total labeled punches, and per-type count with mean punch duration in milliseconds (count\,/\,ms).}
  \label{tab:dataset}
  \centering
  \footnotesize
  \setlength{\tabcolsep}{4pt}
  \resizebox{\textwidth}{!}{%
  \begin{tabular}{l c c r r r r r r r}
      \toprule
      ID & Sex & Ht (m) & Total & Jab & Cross & Lead hook & Rear hook & Lead uppercut & Rear uppercut \\
      \midrule
      E1 & M & 1.89 & 313 & 71\,/\,978 & 50\,/\,984 & 43\,/\,1182 & 47\,/\,959 & 59\,/\,1154 & 43\,/\,955 \\
      E2 & F & 1.80 & 138 & 23\,/\,1126 & 23\,/\,1056 & 25\,/\,1278 & 20\,/\,1252 & 17\,/\,1305 & 30\,/\,1102 \\
      E3 & F & 1.75 & 174 & 34\,/\,1165 & 21\,/\,1395 & 34\,/\,1198 & 24\,/\,1274 & 32\,/\,1263 & 29\,/\,1233 \\
      E4 & M & 1.75 & 151 & 29\,/\,1299 & 21\,/\,1308 & 26\,/\,1174 & 21\,/\,1202 & 23\,/\,1117 & 31\,/\,1168 \\
      E5 & M & 1.75 & 267 & 31\,/\,1787 & 78\,/\,716 & 47\,/\,816 & 38\,/\,760 & 48\,/\,830 & 25\,/\,739 \\
      N1 & F & 1.57 & 499 & 82\,/\,1121 & 100\,/\,1037 & 81\,/\,1309 & 66\,/\,1143 & 85\,/\,1077 & 85\,/\,909 \\
      N2 & M & 1.60 & 296 & 43\,/\,1077 & 47\,/\,762 & 52\,/\,931 & 39\,/\,796 & 58\,/\,804 & 57\,/\,697 \\
      N3 & M & 1.75 & 310 & 70\,/\,901 & 44\,/\,822 & 44\,/\,923 & 51\,/\,885 & 52\,/\,960 & 49\,/\,939 \\
      N4 & F & 1.60 & 222 & 73\,/\,761 & 55\,/\,492 & 25\,/\,837 & 23\,/\,705 & 28\,/\,1122 & 18\,/\,745 \\
      N5 & M & 1.80 & 279 & 55\,/\,1168 & 24\,/\,2255 & 38\,/\,896 & 41\,/\,1522 & 62\,/\,1066 & 59\,/\,776 \\
      N6 & M & 1.75 & 351 & 68\,/\,1244 & 76\,/\,1104 & 53\,/\,1127 & 43\,/\,1264 & 47\,/\,1214 & 64\,/\,1028 \\
      N7 & M & 1.75 & 227 & 46\,/\,1283 & 48\,/\,1122 & 42\,/\,1167 & 36\,/\,1234 & 30\,/\,1228 & 25\,/\,1296 \\
      N8 & M & 1.80 & 418 & 64\,/\,1301 & 88\,/\,1193 & 63\,/\,1034 & 74\,/\,1237 & 54\,/\,1184 & 75\,/\,1080 \\
      N9 & M & 1.60 & 225 & 28\,/\,872 & 42\,/\,765 & 37\,/\,1033 & 27\,/\,795 & 61\,/\,745 & 30\,/\,723 \\
      N10 & F & 1.60 & 274 & 30\,/\,1363 & 61\,/\,1140 & 32\,/\,1197 & 44\,/\,1155 & 48\,/\,1164 & 59\,/\,999 \\
      N11 & F & 1.80 & 172 & 18\,/\,2189 & 28\,/\,2028 & 9\,/\,2379 & 30\,/\,1665 & 72\,/\,863 & 15\,/\,2381 \\
      N12 & M & 1.65 & 397 & 69\,/\,1356 & 58\,/\,949 & 76\,/\,1309 & 70\,/\,1029 & 82\,/\,1245 & 42\,/\,968 \\
      \midrule
      \textbf{All} & 11M/6F & n/a & \textbf{4713} & 834\,/\,1168 & 864\,/\,1041 & 727\,/\,1119 & 694\,/\,1109 & 858\,/\,1055 & 736\,/\,982 \\
      \bottomrule
    \end{tabular}}
\end{table}

%% file: tables/table_per_fold.tex
\begin{table}[t]
  \centering
  \footnotesize
  \setlength{\tabcolsep}{6pt}
  \caption{Per-participant LOPO AUC. Each participant's calibrated scores, averaged over the five seeds, are compared with the opposite-class punches from the other folds. The $\pm$ is the standard deviation of that AUC across the five single-seed heads, and the interval is a bootstrap over participants.}
  \label{tab:perfold}
  \begin{tabular}{lrc}
    \toprule
    Participant & Punches & LOPO AUC \\
    \midrule
    E1 & 313 & $0.503 \pm 0.291$ \\
    E2 & 138 & $0.973 \pm 0.055$ \\
    E3 & 174 & $1.000 \pm 0.002$ \\
    E4 & 151 & $1.000 \pm 0.001$ \\
    E5 & 267 & $0.873 \pm 0.123$ \\
    N1 & 499 & $0.595 \pm 0.136$ \\
    N2 & 296 & $0.763 \pm 0.128$ \\
    N3 & 310 & $0.940 \pm 0.114$ \\
    N4 & 222 & $0.944 \pm 0.068$ \\
    N5 & 279 & $0.877 \pm 0.102$ \\
    N6 & 351 & $0.700 \pm 0.131$ \\
    N7 & 227 & $0.829 \pm 0.118$ \\
    N8 & 418 & $0.962 \pm 0.135$ \\
    N9 & 225 & $0.994 \pm 0.135$ \\
    N10 & 274 & $0.668 \pm 0.173$ \\
    N11 & 172 & $0.851 \pm 0.134$ \\
    N12 & 397 & $0.835 \pm 0.168$ \\
    \midrule
    \multicolumn{2}{l}{\textbf{Mean over participants}} & $\mathbf{0.842} \pm 0.097$ \\
    \multicolumn{2}{l}{\quad bootstrap 95\% CI} & $[0.769, 0.907]$ \\
    \bottomrule
  \end{tabular}
\end{table}

%% file: tables/table_xai.tex
\begin{table}[t]
  \centering
  \footnotesize
  \setlength{\tabcolsep}{6pt}
  \caption{Explanation-quality audit under LOPO. CAA: the mean change in expert
  probability when that joint's or zone's series are permuted across punches,
  split by the held-out participant's class. DCS: change in expert probability
  when a layer's series are substituted with those of the matched expert
  exemplar.}
  \label{tab:xai}
  \begin{tabular}{lcc}
    \toprule
    \multicolumn{3}{l}{\textbf{CAA} --- attribution $|\phi_j|$} \\
    Joint / zone & novice held-out & expert held-out \\
    \midrule
    Pelvis & $0.0207$ & $0.0034$ \\
    L. hip & $0.0191$ & $0.0083$ \\
    R. hip & $0.0057$ & $0.0017$ \\
    Lumbar & $0.0051$ & $0.0044$ \\
    L. shoulder & $0.0042$ & $0.0009$ \\
    R. shoulder & $0.0031$ & $0.0021$ \\
    L. elbow & $0.0067$ & $0.0099$ \\
    R. elbow & $0.0063$ & $0.0031$ \\
    L. heel & $0.0009$ & $0.0002$ \\
    L. midfoot & $0.0008$ & $0.0007$ \\
    L. forefoot & $0.0009$ & $0.0005$ \\
    R. forefoot & $0.0002$ & $0.0001$ \\
    R. midfoot & $0.0004$ & $0.0002$ \\
    R. heel & $0.0004$ & $0.0002$ \\
    \midrule
    \multicolumn{3}{l}{\textbf{DCS} --- counterfactual substitution} \\
    Layer replaced & \multicolumn{2}{c}{mean $\Delta p_k$, novice held-out} \\
    \midrule
    Foundation & \multicolumn{2}{c}{$+0.274$} \\
    Engine & \multicolumn{2}{c}{$+0.271$} \\
    Delivery & \multicolumn{2}{c}{$+0.272$} \\
    \bottomrule
  \end{tabular}
\end{table}

%% file: tables/table_dcs.tex
\begin{table}[t]
  \centering
  \footnotesize
  \setlength{\tabcolsep}{4pt}
  \caption{DCS and CAA per novice participant. Mean rise in expert probability over the participant's punches when each layer is replaced with the matched expert's, the layer with the largest rise (their primary gap), and the joint or zone with the largest CAA attribution on that participant's fold.}
  \label{tab:dcs}
  \begin{tabular}{lcccll}
    \toprule
    Participant & Foundation & Engine & Delivery & Primary gap & Top CAA unit \\
    \midrule
    N1 & $\mathbf{+0.388}$ & $+0.239$ & $+0.282$ & Foundation & L. hip \\
    N2 & $\mathbf{+0.260}$ & $+0.191$ & $+0.162$ & Foundation & L. elbow \\
    N3 & $+0.228$ & $\mathbf{+0.262}$ & $+0.175$ & Engine & L. hip \\
    N4 & $+0.203$ & $+0.309$ & $\mathbf{+0.346}$ & Delivery & Pelvis \\
    N5 & $+0.261$ & $+0.288$ & $\mathbf{+0.334}$ & Delivery & L. hip \\
    N6 & $+0.220$ & $\mathbf{+0.437}$ & $+0.102$ & Engine & Pelvis \\
    N7 & $+0.397$ & $\mathbf{+0.542}$ & $+0.281$ & Engine & R. elbow \\
    N8 & $+0.104$ & $+0.225$ & $\mathbf{+0.465}$ & Delivery & Pelvis \\
    N9 & $\mathbf{+0.624}$ & $+0.117$ & $+0.150$ & Foundation & R. hip \\
    N10 & $+0.087$ & $+0.046$ & $\mathbf{+0.385}$ & Delivery & Pelvis \\
    N11 & $+0.406$ & $\mathbf{+0.419}$ & $+0.341$ & Engine & L. elbow \\
    N12 & $+0.110$ & $+0.181$ & $\mathbf{+0.241}$ & Delivery & L. hip \\
    \midrule
    \textbf{Mean} & $+0.274$ & $+0.271$ & $+0.272$ & & \\
    \bottomrule
  \end{tabular}
\end{table}